# An Application-Specific Design Methodology for STbus Crossbar Generation

Srinivasan Murali, Giovanni De Micheli
Computer Systems Lab
Stanford University
Stanford, California 94305
{smurali, nanni}@stanford.edu

**Abstract**

*As the communication requirements of current and future Multiprocessor Systems on Chips (MPSoCs) continue to increase, scalable communication architectures are needed to support the heavy communication demands of the system. This is reflected in the recent trend that many of the standard bus products such as STbus, have now introduced the capability of designing a crossbar with multiple buses operating in parallel. The crossbar configuration should be designed to closely match the application traffic characteristics and performance requirements. In this work we address this issue of application-specific design of optimal crossbar (using STbus crossbar architecture), satisfying the performance requirements of the application and optimal binding of cores onto the crossbar resources. We present a simulation based design approach that is based on analysis of actual traffic trace of the application, considering local variations in traffic rates, temporal overlap among traffic streams and criticality of traffic streams. Our methodology is applied to several MPSoC designs and the resulting crossbar platforms are validated for performance by cycle-accurate SystemC simulation of the designs. The experimental case studies show large reduction in packet latencies (up to 7×) and large crossbar component savings (up to 3.5×) compared to traditional design approaches.*

## 1 Introduction

As the number of processor/memory cores and the number and size of applications run on *Multiprocessor Systems on Chips (MPSoCs)* increase, the communication between the cores will become a major bottleneck. Traditional communication architectures, such as single shared or bridged buses are inherently non-scalable and will not be able to support the heavy communication traffic [2]. A communication-centric design approach, *Networks on Chips (NoCs)*, has recently emerged as the design paradigm for designing a scalable communication infrastructure for *MPSoCs* [2].

The need for scalable communication architectures is reflected in the recent trend that many of the standard bus products, such as the STbus® (from STMicroelectronics) have now introduced the capability of designing a crossbar with multiple buses operating in parallel, thus providing a low-latency and high-bandwidth communication infrastructure.

**Table 1. Crossbar Performance and Cost**

| Type | Average Lat (in cy) | Maximum Lat (in cy) | Size Ratio |
|---|---|---|---|
| shared | 35.1 | 51 | 1 |
| full | 6 | 9 | 10.5 |
| partial | 9.9 | 20 | 4 |

The communication architecture for the design should closely match the application traffic characteristics and performance requirements. As an example, let us consider a 21-core *MPSoC* running a set of matrix multiplication benchmarks (detailed explanation of the *MPSoC* and experimental set-up is presented in later sections). We consider three different communication architectures using STbus interconnection platform: a shared bus, a full crossbar and a partial crossbar. In Table 1, we present the average and maximum latency incurred by the packets, obtained from SystemC simulation of the design using these platforms and the size of the crossbars (in terms of number of components used) normalized with respect to the size of the shared bus. As seen from the table, as expected, both average and maximum packet latency is much higher for a single shared bus than the partial or full crossbars. However, it is interesting to note that an optimal partial crossbar gives almost the same performance as a full crossbar, even though it uses fewer resources than a full crossbar. A smaller crossbar configuration results in reduction in number of communication components used (such as buses, arbiters, adapters, etc), design area and design power.

In this research we target the design of the optimal STbus crossbar configuration for a given application, satisfying the performance characteristics of the application. The proposed design methodology is based on actual functional traffic analysis of the application, and the generated crossbar configuration is validated by cycle-accurate SystemC simulation of the application using that crossbar. Most previous works on bus generation and NoC topology generation (which are somewhat similar to crossbar generation) are either based on average communication traffic flow between the various cores or based on statistical traffic generating functions. While the former approaches fail to capture local variations in traffic patterns (as the average bandwidth of communication is a single metric that is calculated based on the entire simulation time), the latter approaches are only based on approximations to the functional traffic.

Our design methodology differs from the existing approaches in the fact that, it is based on the analysis of simulated traffic patterns in windows, considering local variations in the communication traffic and reducing the tem-poral overlap among traffic streams mapped onto the same resource. Moreover, our methodology is based on the actual traffic characteristics of the application obtained from SystemC simulations and the designed crossbar configuration is also validated by cycle-accurate simulations. Even though our design approach is fine-tuned for STbus crossbar architecture, it can be easily modified for other crossbar architectures as well. Several experimental case studies on *MPSoC* designs show large reduction in packet latencies (up to $7\times$) and large network component savings (up to $3.5\times$) compared to traditional design approaches.

## 2 Previous Work

A component-based design methodology for SoC design is presented in [1]. The synthesis and instantiation of single bus and multiple bridged buses has been explored in many research works such as [5], [6], [7], [8]. In [9], the authors present an approach for mapping the system's communication requirements and optimizing the communication protocols for a given communication architecture template. In [10], the use of communication architecture tuners to adapt to runtime variability needs of a system is presented.

The need for scalable communication architectures and a communication centric design paradigm, *Networks on Chips (NoCs)*, is presented in [2]. A large body of research such as [12], [13], [11], [16], focus on developing design tools and architectures for NoCs. A detailed survey of many of the NoC research works is presented in [3]. Mapping of communication requirements of a system onto a fixed set of NoC topologies is explored in [20], [21], [14], [15].

In [18] design methodologies for application-specific bus design and in [19], [15], for application-specific NoC topology design are presented. These works are based on average communication transferred between the various cores. In [4], designing application-specific topologies based on actual simulation traces in presented. However, the methodology is based on eliminating contention and can lead to over-sizing of network components, as even a small amount of overlap between two traffic streams would result in the need for separate communication resources for them. In [17], the analysis is based on statistical traffic generators and not functional application traffic.

In this work, our design methodology is based on actual functional traffic of the application. We divide the entire simulation period into a number of fixed-sized windows. Within each window, we guarantee that the application communication requirements (such as the bandwidth requirements) are met. We minimize the overlap among traffic streams mapped onto the same resource, thereby reducing the latency for data transfer. We also consider the criticality and real-time requirements of streams and guarantee that the performance constraints (such as bandwidth and delay constraints) for these streams are met. Our methodology spans an entire design space spectrum with the analysis based on average communication traffic (as done in many previous works) and on peak bandwidth (as done in [4]) being the two extreme design points. Thus our methodology also applies to cases where application traces are not available and only rough estimates of the traffic flows between the various cores is known. The design point in the spectrum is varied by controlling the window size used for the traffic analysis and design, which is explained further in Section 7.2.

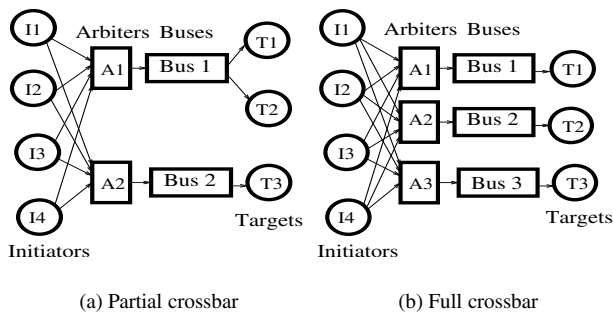

(a) Partial crossbar  (b) Full crossbar

**Figure 1. STbus crossbars**

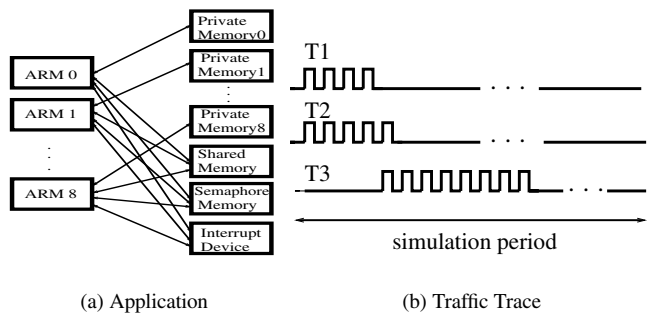

(a) Application  (b) Traffic Trace

**Figure 2. Application Traffic Analysis**

## 3 Problem Definition and Analysis

### 3.1 Problem Definition

The STbus can be instantiated in three ways: as a shared bus, a partial crossbar or a full crossbar. The partial and full crossbars are actually composed of many buses to which the processor/memory cores are connected (refer Figure 1). Two separate STbus crossbars are instantiated for a design: one for the communication from initiators (masters) to targets (slaves) and the other for communication from targets to initiators. The *initiator-target* partial and full crossbars are shown in Figure 1. In this crossbar, all initiators are connected to all buses of the crossbar and one or more targets are connected onto every bus. There are additional interface components: arbiters and frequency/data width adapters (not shown in the figure for clarity) that facilitate the interconnection of heterogeneous processor and memory cores onto the bus. The *target-initiator* crossbar has a similar structure.

The type and size of crossbar needed for an application should closely match the traffic characteristics and performance requirements of the application. A full crossbar, although provides the best performance in terms of minimizing communication latency or maximizing communication throughput, results in a large increase in the number of network components used, design area and power. Note that the size of the two crossbars (the *initiator-target* and the *target-initiator*) can be different.

### 3.2 Application Traffic Analysis

In this subsection, we explore the traffic characteristics of applications and formulate the performance constraints to be satisfied by the crossbar designed for the system. As an example, we consider the 21-core matrix multiplication application shown in Figure 2(a). In this example, there are 9 `ARM` cores, their private memories, a shared memory for inter-processor communication, a semaphore memory for maintaining locks for shared memory accesses and an interrupt device. The `ARM` cores act as initiators and the memory cores act as targets. The `ARM` cores run a set of pipelined matrix multiplication benchmarks that involve accesses to their private memory and inter-processor communication through the shared memory. We performed a cycle-accurate simulation of the system with full `STbus` *initiator-target* and *target-initiator* crossbars. A small trace of the traffic to three of the targets is shown in Figure 2(b).

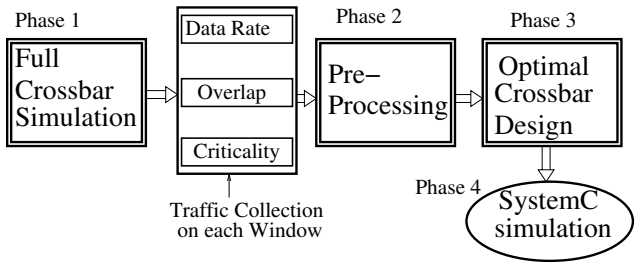

**Figure 3. Design Methodology**

Even though the aggregate traffic (measured over the entire simulation period) to the three targets is lower than that can be supported by a single bus, using a single bus to connect all three targets will lead to high average and peak latency due to overlap in traffic patterns during some regions of the simulation. Another related point is that if overlap are not considered, connecting targets 1 and 2 on to the same bus is better than connecting targets 1 and 3 onto the same bus, as the former results in lower bandwidth needs. However, the latter solution will result in better performance (reduced latency) while still satisfying the bandwidth needs. Note that using peak bandwidth instead of the average bandwidth will solve this problem, but lead to an over-design of the crossbar (in terms of number of buses needed or their frequency of operation). The design methodology needs to consider overlap among the various traffic streams into account and should consider local variations in traffic rates. Moreover, the designed crossbar should be such that, while minimizing the average latency, should also minimize the maximum latency that any packet or traffic stream can incur. Also, some of the traffic streams can be more critical than the others and such real-time traffic streams need to be given guaranteed real-time performance.

## 4  Design Methodology

The design flow for the crossbar design is shown in Figure 3. The application is initially designed using full crossbar for *initiator-target* and *target-initiator* communication and a SystemC simulation of the design is carried out. For the simulations, we use the `MPARM` simulation environment [22] that allows interconnection of `ARM` cores to several interconnection platforms (such as `AMBA`, `STbus`, ...) and to perform cycle accurate simulations for a variety of benchmark applications. We present here only the design of the `initiator-target` crossbar, as the `target-initiator` crossbar can be designed in a similar fashion.

To effectively capture local variations in traffic patterns and to perform overlap calculations, we define a window-based traffic analysis. The entire simulation period is divided into a number of windows and the traffic characteristics to the various targets in each window are obtained. The traffic characteristics recorded include: the amount of data received by each target in every window, amount of pair-wise overlap between the traffic streams to the targets in every window, the real-time requirements of traffic streams, etc. Without loss of generality, in the rest of the paper we assume that all the windows are of equal size, although the methodology also applies to windows with varying sizes. The size of the window is parameterizable and depends on the application characteristics and performance requirements. The effect of the window size on the quality of the solution is explored in Section 7.2.

After the data collection phase, a pre-processing phase is carried out in which the targets that have large overlap in any window and need to be put on different buses are identified. In this phase, the overlapping critical streams that need to be on separate buses are also identified. The maximum number of targets that can be connected to a single bus, to bound the maximum latency, is also identified in this phase.

In the next phase, the optimal crossbar configuration for the application, satisfying the performance constraints is obtained. To generate the optimal crossbar configuration, we use the traffic information collected in each window and check whether the bandwidth, overlap and criticality constraints are satisfied in each window.

In the final phase, the resulting crossbars are instantiated in the `MPARM` environment and SystemC simulations are carried out.

## 5  Problem Formulation

In this section we formulate the mathematical models of the crossbar design problem.

**Definition 1** *The set of all targets is represented by the set $T$. The set of all windows used for traffic analysis is represented by the set $W$, with the length of each window (in terms of number of cycles) represented by $WS$. The set of buses used in the crossbar is represented by the set $B$.*

**Definition 2** *The number of cycles that each target $t_i$, $\forall i \in 1..|T|$, receives data in every window $m$, $\forall m \in 1..|W|$, is represented by $comm_{i,m}$ [1]. The amount of data overlap (in number of cycles) between every pair of targets ($t_i$, $t_j$) in each window $m$ is represented by $wo_{i,j,m}$.*

The overlap between every pair of targets $t_i$ and $t_j$, over the entire simulation period is obtained by summing the overlap between them in all the windows and represented by the entries of the overlap matrix $OM$:

$$om_{i,j} = \sum_m wo_{i,j,m} \quad : \forall i, j \quad (1)$$

In the pre-processing step of the design flow (refer Figure 3), those pair of targets that have overlap exceeding the threshold value (which is parameterizable) in any window are identified. By placing such targets onto separate buses, the maximum and average latency of data transmission can be reduced and in some cases can also speed up the process of finding the optimal crossbar configuration. The effect of this pre-processing step is explored in detail in Section 7.3. Also in this pre-processing step, the real-time traffic streams that overlap with each other in any window are identified. Such targets with overlapping real-time streams should not be placed on the same bus as real-time communication guarantee to the streams cannot be given in this case. We define the set of all targets that cannot be on the same bus by the conflict matrix:

$$c_{i,j} = \begin{cases} 1, & \text{if } t_i \text{ \& } t_j \text{ should be on different buses} \\ 0, & \text{otherwise} \end{cases} : \forall i,j \quad (2)$$

---

[1] In the rest of this paper we follow the convention that variables $i$ and $j$ are defined for $1..|T|$, variable $k$ is defined for $1..|B|$ and $m$ for $1..|W|$.

We model the performance constraints that need to be satisfied by the crossbar configuration in each window as constraints of a Mixed Integer Linear Program (MILP).

**Definition 3** *The set $X$ represents the set of binding variables $x_{i,k}$, such that $x_{i,k}$ is one when target $t_i$ is connected to the bus $b_k$ and zero otherwise.*

In the STbus crossbar, each target has to be connected to a single bus (while a single bus can connect multiple targets, as shown in Figure 1(a)). This is implemented by the following constraint:

$$\sum_k x_{i,k} = 1 \quad : \forall i \quad (3)$$

In every window of the traffic analysis, the individual buses in the crossbar have to support the traffic through them in that window. By evaluating the bandwidth constraints over a smaller sample space of a window (which is typically few thousand cycles) instead of the entire simulation sample space (which can be millions of cycles) we are better able to track the local variations in the traffic characteristics. This window-based bandwidth constraint is represented by the equation:

$$\sum_i comm_{i,m} \times x_{i,k} \leq WS \quad : \forall k,m \quad (4)$$

**Definition 4** *The set $SB$ represents the set of sharing variables $sb_{i,j,k}$, such that $sb_{i,j,k}$ is one when targets $t_i$ and $t_j$ share the same bus $b_k$ and zero otherwise. The set $S$ represents the set of sharing variables $s_{i,j}$, such that $s_{i,j}$ is one when targets $t_i$ and $t_j$ share any of the buses of the crossbar and zero otherwise.*

The $sb_{i,j,k}$ can be computed as a product of $x_{i,k}$ and $x_{j,k}$. However, this results in non-linear (quadratic) equality constraints. To break the quadratic equalities into linear inequalities, we use the following set of equations:

$$sb_{i,j,k} \in \{0,1\}$$
$$x_{i,k} + x_{j,k} - 1 \leq sb_{i,j,k}$$
$$0.5\, x_{i,k} + 0.5\, x_{j,k} \geq sb_{i,j,k} \quad : \forall i,\ j,\ k \quad (5)$$

and the $s_{i,k}$ are computed using the equation:

$$s_{i,j} = \sum_k sb_{i,j,k} \quad : \forall i,j \quad (6)$$

The condition that certain targets are forbidden to be on the same bus, obtained from Equation 2, is represented by:

$$c_{i,j} \times s_{i,j} = 0 \quad : \forall i,j \quad (7)$$

As the number of targets onto a single bus increases, even for traffic streams that don't have substantial overlap, the maximum latency that can be incurred by a packet/traffic stream can increase substantially. In the worst case, packets to all the targets onto a bus can arrive in the same cycle, which need to be serialized on the bus, thereby making the maximum latency incurred by packets to some of the targets higher. In order to reduce the maximum delay that can be incurred by any packet, we can restrict the maximum number of targets that can be connected to the same bus below a threshold ($maxtb$) that is parameterizable. This is represented by the following constraint:

$$\sum_i x_{i,k} \leq maxtb \quad : \forall k \quad (8)$$

The fact that all the integer variables introduced above take values of either 0 or 1 only, is represented by:

$$x_{i,k}, s_{i,j}, c_{i,j} \in \{0,1\} \quad : \forall i,j,k \quad (9)$$

## 6 Crossbar Design Algorithm

The algorithm for the STbus crossbar design has two major steps: the first is to find the minimum crossbar configuration that satisfies the performance constraints (that were presented in the above section) and the second step is to find the optimal binding of the targets to the chosen crossbar configuration.

In order to find the best crossbar configuration, all possible configurations are tested in a binary search manner to find the minimum configuration that satisfies the performance constraints that were modeled as MILP constraints in the previous section.

The following MILP is tested for a feasible solution for each configuration until the best configuration is obtained:

$$\text{obj: Feasibility Analysis}$$
$$\text{subject to Equations (3) to (9).} \quad (10)$$

Note that the MILP has no objective function as the aim is to just test for feasibility.

Once the best crossbar configuration is obtained, in the next step, the optimal binding of the targets onto buses of the crossbar is obtained. A binding of targets to the buses that minimizes the amount of overlap of traffic on each bus will result in lower average and peak latency for data transfer. For this, the above MILP is solved with the objective of reducing the maximum overlap on each of the bus, and satisfying the performance constraints, as follows:

$$\text{min: } maxov$$
$$\text{s.t. } \sum_i \sum_j om_{i,j} \times sb_{i,j,k} \leq maxov \quad : \forall k$$
$$\text{and subject to Equations (3) to (9).} \quad (11)$$

By splitting the problem into two MILPs, we speed up the execution time of the algorithm as solving MILP1 for feasibility check is usually faster than solving the MILP2 with objective function and additional constraints. The MILPs are solved using the CPLEX package [23]. The runtime of the algorithm for all our simulation studies was under few hours, when run on a 1GHz SUN machine. The runtime for the MILP is not that large as the largest possible STbus crossbar size (and maximum number of targets) is 32 and hence the number of integer variables is less than few thousand.

## 7 Experiments and Case Studies

### 7.1 Application Benchmark Analysis

We applied our design methodology for crossbar design of several *MPSoCs*: *Matrix suite-1 (Mat1-*25 *cores), Matrix suite-2 (Mat2-*21 *cores), FFT suite (FFT-*29 *cores), Quick Sort suite (QSort-*15 *cores) and DES encryption system (DES-*19 *cores)*. For comparison purposes, we also designed crossbars based on average communication traffic flows (as done in previous approaches), by relaxing overlap constraints and using a single window for analysis. The SystemC simulation of the applications were carried out on the STbus platform using the designed crossbars. We briefly analyze here the quality of the crossbar design obtained for the matrix multiplication benchmark, *Mat2* (refer Figure 2). In this benchmark, there are 9 initiators and 12 targets. Of the 12 targets, accesses to 3 of the targets (the shared memory, semaphore memory and interrupt device) is much lower than the accesses to private memories, as these are only used for inter-processor communication. There is substantial temporal overlap between the traffic flows from the various ARM cores to their private memories, as the ARM cores perform similar computations and thus access their memories at almost the same time. In order to satisfy the window bandwidth constraints, only few of them can share a single bus. Our methodology, when applied to this benchmark, results in the use of 3 buses for the initiator-target crossbar. Each of the bus has 3 private memories and one of the common memories connected to it. Moreover, the bindings are such that, the targets with highly overlapping streams are placed on different buses, an important design constraint explained in detail in Section 3.2. As a result, the designed crossbar has acceptable performance (in terms of average and maximum latency constraints) with $3.5\times$ reduction in the number of buses used, when compared to a full crossbar.

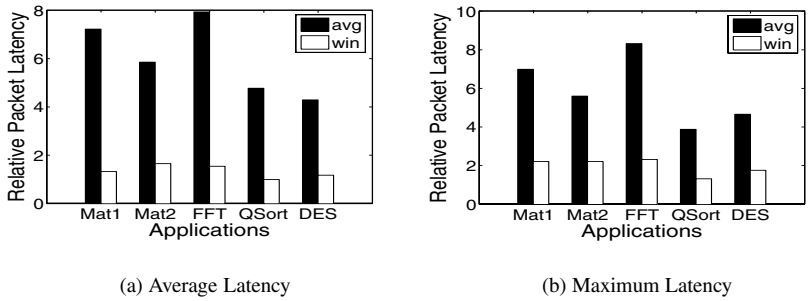

(a) Average Latency  (b) Maximum Latency

**Table 2. component savings**

| Application | Full crossbar bus count | Designed crossbar bus count | Ratio |
|---|---|---|---|
| Mat1 | 25 | 8 | 3.13 |
| Mat2 | 21 | 6 | 3.5 |
| FFT | 29 | 15 | 1.93 |
| QSort | 15 | 6 | 2.5 |
| DES | 19 | 6 | 3.12 |

**Figure 4. Application relative latencies**

The average and maximum packet latencies for the applications obtained from the simulations, normalized with respect to the latencies incurred in a full crossbar system, are presented in Figures 4(a) and 4(b). As seen from the figure, the latencies incurred by crossbar designs based on average traffic flows are $4\times$ to $7\times$ higher than the crossbars designed using our scheme. Also, the latencies incurred in the designs generated by our scheme are within acceptable bounds from the minimum possible latencies (of a full crossbar). Moreover, depending on the design objective, crossbar size-performance trade-offs can be explored in our approach by tuning the analysis parameters (such as the window size, overlap threshold, etc.), as explained in the next subsections. In Table 2, we compare the number of buses used in the crossbar designed by our approach with that of full crossbar for the applications. We get large reduction (up to $3.5\times$) in the crossbar size by using our design approach.

### 7.2 Window Sizing

The size of the window used during the design process is an important parameter that determines the efficiency of the design methodology to capture the application performance parameters. A small window size results in much finer control of the application performance parameters and the resulting crossbars have lower latencies. However, a very small window size will lead to over-design of the network components. On the other hand, a large window size results in lesser control over the performance parameters of the application, but results in a more conservative design approach where higher packet latencies can be tolerated.

To illustrate these effects, we applied our design methodology with different window sizes for a synthetic benchmark with 20 cores. The typical burst sizes (we refer to a burst as a stream of packets generated by the same core) for the benchmark were around 1000 cycles. When the window size is much smaller than the burst size, the size of the crossbar generated is very close to that of a full crossbar (refer Figure 5(a)). When the window size is around few times that of the burst size (from 1-4 times), crossbar designed by our approach has much smaller size (typically around 25%) and acceptable latencies (around $1.5\times$) of that of a full crossbar. For aggressive designs, the window size can be set closer to the burst size and for conservative designs (where larger packet latencies can be tolerated), the window size can be set to few times the typical burst size. The acceptable window sizes for various burst sizes is presented in Figure 5(b). It can be seen from the plot that the window size varies almost linearly with the burst size, consolidating the above arguments.

### 7.3 Real-Time Streams & Effect of Binding

In each simulation window, the critical traffic streams that require real-time guarantees are recorded. During the pre-processing step of the design flow (refer Figure 3), the real-time traffic streams that overlap with each other in any window are identified. In order to provide real-time guarantees to such streams, in our methodology the cores with the overlapping critical streams are placed on separate buses of the crossbar. Experimental results on the benchmark applications show a very low packet latency (almost equal to the latency of perfect communication using a full crossbar) for such streams.

After finding the best crossbar configuration, we do an optimal binding of the cores onto the buses of the crossbar, minimizing the total overlap on each bus. By minimizing the overlap on each bus, the packet latencies reduce significantly. To illustrate this effect, we compare the crossbars designed using our approach with two binding schemes: random binding of cores onto the buses, satisfying the design constraints (Equations 3-9) and optimal binding that

Proceedings of the Design, Automation and Test in Europe Conference and Exhibition (DATE'05)
1530-1591/05 $ 20.00 IEEE

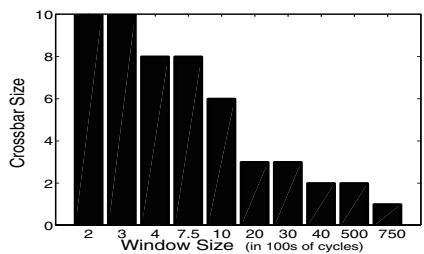
(a) Initiator-Target crossbar vs. window size

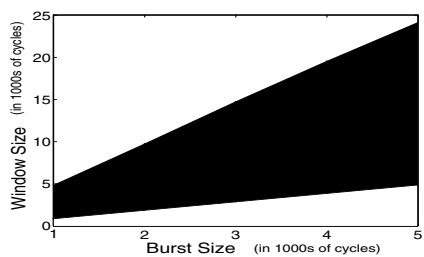
(b) Burst vs. Window size

**Figure 5. Effect of window size variations**

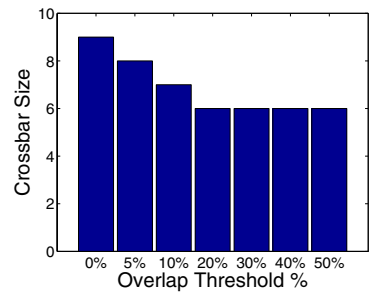
**Figure 6. Overlap threshold effects**

minimizes overlap on each bus, satisfying the design constraints. The average latency incurred by the random binding scheme for the benchmark applications was on an average $2.1\times$ higher than that incurred by the optimal binding scheme.

### 7.4 Overlap Threshold Setting

By varying the three parameters: window size, overlap threshold and maximum number of targets (or initiators) on a bus, the crossbar can be designed such that the average and, more so, the maximum packet latencies incurred in the design are acceptable. The effect of the overlap threshold parameter on the size of the `initiator-target` crossbar generated for the synthetic benchmark is presented in Figure 6. The plots end at 50% overlap between targets because, if the pair-wise overlap between two targets exceeds $50\%$ of the window size (in any of the windows), then the window bandwidth constraints cannot be satisfied. So, the maximum value of the overlap parameter can be set at 50% of the slot size. This will also speed-up the process of finding the best crossbar configuration, as such overlapping targets will be identified in the pre-processing phase (refer Figure 3) and will be forbidden to be on the same bus of the crossbar. From experiments, we found that for aggressive designs (where there are tight requirements on the maximum latencies) the threshold can be set to around 10% and for conservative designs, the threshold can be set to 30%-40% of the window size.

## 8 Conclusions and Future Work

To accommodate the growing communication demands of *Multiprocessor Systems on Chips (MPSoCs)*, scalable communication architectures and related design methodologies are needed. In this work, we have presented a design methodology for designing the optimal crossbar configuration for an application and for binding the cores onto the crossbar resources. Our design approach is based on a simulation window-based analysis of the application traces, considering the local variations in traffic rates, temporal overlap among traffic patterns and criticality of traffic streams. The methodology has several parameters that can be tuned to explore the design space of the crossbar design and to match the application characteristics. The design methodology, though fine-tuned to STbus crossbar architecture, can be easily extended to other crossbar architectures as well. Several experimental studies show large reduction in latency and crossbar components usage, compared to traditional design approaches. In future, we plan to analyze the effect of using variable simulation window sizes for the design for guaranteeing *Quality-of-Service (QoS)* for applications.

## 9 Acknowledgements

This research is supported by MARCO Gigascale Systems Research Center (GSRC) and the US National Science Foundation (under contract CCR-0305718).